# Clusters of microRNAs emerge by new hairpins in existing transcripts

Antonio Marco[*], Maria Ninova, Matthew Ronshaugen, Sam Griffiths-Jones

Faculty of Life Sciences, Michael Smith Building, Oxford Road,

University of Manchester, Manchester, M13 9PT, United Kingdom

[*] Corresponding author:

E-mail: amarco.bio@gmail.com

Tel. +44 (0) 161 275 1565

# ABSTRACT

Genetic linkage may result in the expression of multiple products from a polycistronic transcript, under the control of a single promoter. In animals, protein-coding polycistronic transcripts are rare. However, microRNAs are frequently clustered in the genomes of animals, and these clusters are often transcribed as a single unit. The evolution of microRNA clusters has been the subject of much speculation, and a selective advantage of clusters of functionally related microRNAs is often proposed. However, the origin of microRNA clusters has not been so far explored. Here we study the evolution of microRNA clusters in *Drosophila melanogaster*. We observed that the majority of microRNA clusters arose by the *de novo* formation of new microRNA-like hairpins in existing microRNA transcripts. Some clusters also emerged by tandem duplication of a single microRNA. Comparative genomics show that these clusters are unlikely to split or undergo rearrangements. We did not find any instances of clusters appearing by rearrangement of pre-existing microRNA genes. We propose a model for microRNA cluster evolution in which selection over one of the microRNAs in the cluster interferes with the evolution of the other linked microRNAs. Our analysis suggests that the study of microRNAs and small RNAs must consider linkage associations.

# INTRODUCTION

MicroRNAs are small endogenous RNA sequences involved in the regulation of essentially all biological processes in animals and plants (1–3). MicroRNAs are produced from longer transcripts by the RNA interference machinery (reviewed in [4, 5]). A striking feature of these molecules is that their loci are often clustered in the genome (6–8). According to miRBase (9), more than 30% of animal microRNAs are organized into clusters, some of which have been experimentally shown to produce polycistronic transcripts (10–12). Hence, multiple microRNAs can be produced from the same primary transcript. Further studies including microRNA co-expression and primary transcript identification suggest that the majority of microRNA clusters are indeed transcribed as a single unit (13–16).

The evolutionary importance of microRNA clusters has been the subject of much speculation. Many clusters contain members of the same family, suggesting an important role of gene duplication in their evolution (17, 18). However, clusters very often contain members of different microRNA families, particularly in animal genomes (reviewed in [1]). Since co-transcription is often used to imply a functional relationship, unrelated microRNAs in the same cluster are often assumed to have similar targeting properties, for example targeting genes in the same pathway (19). However, the origin and evolution of microRNA clusters has not been investigated in detail.

There are a number of known types of polycistronic transcripts, each of which suggests a possible mode of evolution for polycistronic microRNAs. Bacterial operons are formed by multiple protein coding loci under the control of a single promoter. These loci are transcribed as a single transcriptional unit and then the different open reading frames are translated separately by the ribosome (20). The evolutionary origin of bacterial operons has been extensively debated, and several models of evolution have been proposed (21). A common feature of the many models is that genes in the same operon are functionally related, that is, participate in the same biochemical pathway (21, 22). We define this general model as the "*put together*" model, which suggests that functionally related products become regulated under a common promoter during evolution (Figure 1A). Under this hypothesis, evolutionarily unrelated microRNAs scattered around the genome may become clustered together during evolution. This mode of evolution has been suggested to explain the existence of clusters of microRNAs from different families (19). Operons have also been found in some animals, particularly in the nematode *Caenorhabditis elegans* (23) and the ascidian *Ciona intestinalis* (24). Operon formation in nematodes is found to be a one-way phenomenon due to molecular constraints (23). Comparative genomics analysis of *C. elegans* and related species reveals that their operons appeared as a by-product of genome reduction, leaving unrelated genes under the control of a single promoter (25, 26). We define this mechanism as the "*left together*" model (Figure 1B), under which microRNAs would be organized into clusters as a stochastic by-product of genome reorganization. More recently, polycistronic transcripts encoding small peptides have been found in arthropods (27). For example, the gene *mille-pattes* is an essential gene during early development, and codes for a number of small peptides (27). Since these peptides are very similar in sequence, an origin of polycistronic transcription by tandem gene duplication is plausible. MicroRNA cluster formation by gene duplication has been observed in animals (17) and probably dominates the evolution of plant microRNA clusters (18, 28). This is the "*tandem duplication*" model (Figure 1C).

However, a fourth mechanism of cluster formation is possible in the case of microRNAs. Any transcript with a hairpin structure is potentially a target of the RNases Drosha and Dicer. The cleavage of a precursor microRNA is largely independent of its specific nucleotide sequence (29, 30). Thus, many transcribed hairpins in the genome are potential targets of Drosha and Dicer. Indeed, microRNAs arise *de novo* in the genome at a high rate (31, 32). Hence, it is plausible that the emergence of a new hairpin near to an existing microRNA could lead to formation of a microRNA cluster, as has been suggested for the vertebrate mir-17 cluster, for example (33). We call this the "*new hairpin*" model (Figure 1D).

The evolutionary origin of microRNA clusters has not been systematically studied. We explore in this paper the source of all *Drosophila melanogaster* clusters by tracing the evolution of their microRNAs, and evaluate the relative contribution of the different microRNA cluster formation models.

## MATERIALS AND METHODS

MicroRNA sequences, genomic coordinates and expression datasets for *D. melanogaster* were extracted from miRBase version 18 (9). We define a cluster of microRNAs as a group of microRNA precursors with an inter-microRNA distance of less than 10kb on the same genomic strand. The degree of co-expression of clustered microRNAs was calculated as the Pearson correlation coefficient of the absolute read counts between all tissues/developmental stages from available RNAseq experiments. We compile homologous microRNAs in animals from miRBase microRNA family annotation, and from BLAST searches (34) with parameters: w=4, r=2, q=-3, against multiple genome sequences (Supplementary Table 1). We also included in our analysis the microRNA families described by Sempere, Wheeler and collaborators (35, 36). We aligned sequences with Clustal X 2.0 (37) and MAFFT 6.85 (38), manually refined the alignments with RALEE (39), and reconstructed evolutionary trees with standard phylogenetic methods: neighbor-joining (40) and maximum likelihood (41), using MEGA5 (42).

To determine the evolutionary origin of each cluster, we first determined the age of each of the microRNAs in the cluster by analysing sequence alignments and phylogenetic trees of microRNA families (Supplementary Dataset 2). We then identified the two original (oldest) microRNAs, and examined the nature of the event that led to these two microRNAs to be clustered together. If the two oldest members of a cluster belong to the same microRNA family, we inferred that the cluster emerged by tandem duplication (Figure 1C). Otherwise, the cluster was formed by one of the other models (Figure 1A,B and D). If the two original microRNAs derive from disparate loci in any other genome, the cluster may have originated by a fusion event. Otherwise, if the two original microRNAs always appear together, we conclude that the cluster was formed by *de novo* emergence of a novel microRNA family. Multiple sequence alignments of related microRNAs are available in the supporting information Supplementary Dataset 2. MicroRNA expression datasets are detailed in Supplementary Table 2.

**RESULTS**

**MicroRNA clusters in *Drosophila melanogaster***

We have studied the genomic distribution and evolutionary origin of 238 *D. melanogaster* microRNAs (see Materials and Methods). These microRNAs are highly clustered in the genome, with 74 (31%) of the annotated sequences less than 10 kb away from another microRNA. Analysis of expression data from different tissues/developmental stages shows that, on average, microRNAs separated by less than 10 kb are highly co-expressed (Figure 2A). The median distance between two clustered microRNAs is only about 130 nucleotides, indicating that clustered microRNAs are, in general, tightly linked in the genome. This observation is in agreement with previous analysis on a more limited dataset (43) and supports 10 kb as an appropriate global threshold for defining clusters of microRNAs that are co-expressed. These clusters are most likely produced from single primary transcripts under the control of a single promoter (16). Using this criterion, we defined 21 *Drosophila* microRNA clusters (Table 1).

The number of microRNAs in each cluster is variable, although the majority are small: of size 2-3 (Figure 2B; white boxes). The distribution of the number of different microRNA families in the same cluster shows that only 4 out of the 21 clusters are formed by a single family (Figure 2B; black boxes). We plotted the size of each cluster against the number of families, and observed that clusters of sizes 2 and 3 (the most abundant; Figure 2B) are more likely to be composed of members of different microRNA families (Figure 2C). This suggests that the initial microRNA cluster-forming event is rarely tandem duplication (Figure 1C), and alternative models should be considered (Figure 1).

**Evolutionary origin of MicroRNA clusters**

We reconstructed the evolutionary origin of all *D. melanogaster* microRNA clusters by phylogenetic analyses of their members and prediction of homologous microRNAs in other animal species (see Materials and Methods). A summary of the 21 identified clusters is shown in Table 1, and a more detailed analysis in the Supplementary Dataset 1. Seven clusters (33%) are specific to drosophilids (Table 1, Figure 3). Collectively, 14 clusters (the majority of our dataset) emerged within the insects (Figure 3), that is, the Melanogaster, *Drosophila*, and insect lineages in Table 1. Two clusters are conserved among all metazoans: the mir-125/let-7/mir-100 and the mir-92a/mir-92b clusters.

We can find no cases where clustered microRNAs in *D. melanogaster* have homologs that derive from disparate loci in any other genome. We therefore conclude that none of the *D. melanogaster* clusters emerged by the union of pre-existing single microRNAs. This rules out two of our evolutionary models of cluster origin: *put together* and *left together*. The initial cluster-forming events for all extant microRNA clusters are predicted to be tandem duplication and hairpin formation (Figure 3), with the latter being the most common (13 out of the 21 cases). The seven new clusters that emerged in the last common ancestor of drosophilids are conserved in all extant (studied) species, supporting the notion that these clusters are evolutionarily constrained after their emergence (Figure 3). Around 15% (14/99) of the microRNAs that emerged *de novo* in the Melanogaster lineage are clustered with another microRNA. However, more than 50% (35/66) of the microRNAs that emerged *de novo* prior to the split of the *Drosophila* lineage are clustered. As we look at sets of microRNAs of increasing age, the proportion that have arisen by *de novo* hairpin formation quickly approaches the 30% of observed clustered microRNAs in most species. This indicates that microRNAs in clusters are less likely to be lost after they emerge than non-clustered microRNAs. We conclude that microRNA clusters in *D. melanogaster* primarily originated by *de novo* hairpin formation.

**MicroRNA clusters are evolutionarily stable to genomic reorganizations**

A fraction of the microRNAs that emerged within the dipteran lineage are less than 10 kb apart from another microRNA (62 out of 178). We therefore speculate that clusters are important generators of microRNAs that may later become independent transcripts by translocation or duplication out of the original cluster. Thus, we explored whether extant non-clustered *D. melanogaster* microRNAs are clustered in any other animal genome, by systematic search for potential microRNA homologs of *Drosophila* microRNAs in other species (see Materials and Methods). On first inspection, it does indeed appear that *Drosophila* non-clustered microRNAs have clustered homologs in other species (Table 2). However, close examination of this dataset reveals that the majority of these clusters were the product of independent local tandem duplication or new hairpin formation. For instance, in mammalian genomes mir-7 is clustered with mir-1179, a mammal-specific microRNA, showing that the creation of new clusters by new hairpin formation also happens in other clades (Table 2). Similarly, mir-285 has been tandemly duplicated in the vertebrate lineage (Table 2).

We have found two instances of microRNA clusters in animals whose individual microRNAs are apparently not clustered in *Drosophila* (mir-1/mir-133 and mir-276a/mir-276b; Table 2). However, both pairs of microRNAs are also linked in the *Drosophila* genome, although with an inter-microRNA distance of greater than 10 kb (see also [44]), thereby escaping our conservative cluster definition. There are two further cases of *Drosophila* non-clustered microRNAs that are clustered in another organism. First, mir-87 forms a cluster of two duplicates in most studied animals yet *Drosophila* conserves only a single copy. This may be a rare case of "acquired individuality" by loss of one of the microRNAs in a cluster. The other case is mir-276a/b. These two microRNAs are not clustered in any species except in the crustacean *Daphnia pulex*. The most likely explanation is that mir-276a/b in *Daphnia* resulted from an independent, lineage-specific, gene duplication. We also observed that mir-9 and mir-279 microRNAs appear clustered in some insects (*Apis mellifera* and *Tribolium castaneum* according to miRBase) suggesting that an original cluster may have split in *Drosophila*. However, the evolution of the mir-9 family is particularly complex and will be better understood as new genome sequences become available. In summary, clusters of microRNAs are evolutionary units that are rarely the source of singleton microRNAs. In most cases, after a cluster is formed in the genome, it either stays together or it is lost as a whole.

## DISCUSSION

In this work we have investigated the evolutionary origin of microRNA clusters studying the model organism *D. melanogaster*. Contrary to observations in other types of polycistronic transcripts, microRNA clusters mostly emerged by tandem duplication and *de novo* hairpin formation in existing microRNA transcripts, with the latter being the dominant mechanism. Only two clusters are conserved in all metazoans, mir-92a/mir-92b and mir-125/let-7/mir-100. However, mir-92a/mir-92b may be the product of independent duplications in different animal lineages, that is, mir-92a/mir-92b of protostomes and deutoerostomes may not be orthologous clusters (Supplementary Figure 1). Although the statistical support of our phylogenetic analysis is weak (low bootstrap values) the fact that there is only one copy in *Daphnia pulex* also supports an insect specific duplication of mir-92. Moreover, mir-92a in *Drosophila* is hosted inside an intron whilst mir-92b is not, suggesting that the two microRNAs may not be part of the same transcript. The other cluster, mir-125/let-7/mir-100, is probably the only conserved cluster in most metazoans. Indeed, mir-100 is the evolutionarily most ancient microRNA and it is conserved in metazoans and cnidarians (45, 46).

Tandem duplication has been described as an important source of polycistronic microRNAs in plants (18, 47) and in animals (17). Our analysis supports the view that this mechanism is more important in the formation of clusters in plants (3, 47) as we find only five cases in which a tandem duplication is the original microRNA cluster-forming event (Table 1). The remaining clustered duplicates arose after the cluster-forming event. Two of the five clusters, mir-13b-1/mir-13a/mir-2c and mir-2a-2/mir-2a-1/mir-2b-2, are derived from a single ancestral mir-2/mir-13 cluster (48, 49). All members of the mir-2/mir-13 ancestral cluster belong to the same family (the mir-2 family), suggesting that the ancestral cluster originated by tandem duplication. However, we have previously shown that the mir-2 cluster originally appeared by the *de novo* birth of the first mir-2 family member within the mir-71 transcript (48, 49). Later, the mir-2 family expanded by duplication and mir-71 was lost in several lineages, including the *Drosophila* genus (49). This example shows that cluster formation by novel acquisition of a hairpin may be masked by subsequent microRNA gene loss. Hence, our approach is likely to underestimate the number of clusters formed by novel hairpin formation. Another caveat is that the actual age of some clusters may be greater than we detect with our conservative methodology. Ongoing work in our lab suggests, for instance, that the mir-6-3~mir-309 cluster may be conserved beyond dipterans (Ninova, Ronshaugen and Griffiths-Jones; in preparation).

Tandem duplication is important in the evolution of already existing clusters, and may generate novel functions of existing microRNAs (43). With the available data, we can only speculate why duplication is much less frequent in cluster formation in animals than in plants. Plant microRNAs frequently target gene transcripts with high complementarity, whilst animal microRNAs bind their targets with more mismatches (50). Two tandemly duplicated microRNAs could therefore quickly diversify in their targeting properties in plants, whereas it may take longer to accumulate sufficient changes in animals to modify their targets. Tandemly duplicated microRNAs in animals are therefore more likely to be functionally redundant in the long term. For instance, members of the mir-2 family have, in general, the same targets (51, 52, 49). In addition, an animal microRNA duplicated in tandem may produce a gene dosage imbalance. However, the emergence of a new microRNA in an existing microRNA transcript will not affect the existing regulatory network. Protein-coding genes tend to diversify their expression pattern after duplication (53). However, duplicated microRNAs encoded in the same transcript may not be able to diversify unless they break the linkage. Some authors have suggested that, since plant microRNAs have high complementarity to their targets, it is less likely that novel microRNAs acquire functional targets in plants, explaining why *de novo* emergence is less important than duplication important in these species (see discussion in [47]). However, this explanation assumes that a new microRNA will have functional targets as soon as it emerges in the genome. Our analyses indicate that that may not be always true, as linkage associations could play an important role in the fixation of new microRNAs. Further analyses of the increasing amount of plant microRNA datasets will clarify the evolutionary fate of novel microRNAs in plants.

Our data show that clusters of microRNAs generally evolve as single units and are lost as a whole, probably because of the tight linkage of the microRNAs. This cluster stability is known for nematode gene clusters as well (25, 54), where cluster (operon) formation is described as a "one-way" evolutionary process (23). Our comparative genomics exploration of animal microRNAs also indicates that microRNA clusters often gain new microRNAs (either by tandem duplication of further new hairpin acquisitions) yet they rarely split or suffer rearrangements. In principle, microRNA hairpins can arise randomly in any genomic position. However, new hairpins within microRNA encoding transcripts may be more likely to become functional microRNAs, as these transcripts are already interacting with the small RNA processing machinery. Indeed, it has been found recently that primary microRNA transcripts include various sequence motifs that are required for the proper processing of precursor microRNAs (55). Clustered microRNAs are actually very close to each other (median distance of 130 nucleotides in our study) suggesting that any regulatory motif in the primary transcript may affect all the microRNAs in the cluster. MicroRNAs can also be lost from existing clusters, although this is relatively infrequent. A notable case is the mir-125/let-7/mir-100 cluster, which is highly conserved across the animal kingdom, although in both Nematodes (56) and in Platyhelminthes (57) mir-125 and let-7 are not clustered, and mir-100 is lost. This exceptional case shows that highly conserved linkage associations between microRNAs can be lost during evolution without major consequences.

Recombination between two closely linked loci by crossing-over is very unlikely. Consequently, selection operating on one microRNA in a cluster results in greatly reduced selection efficiency in the neighbouring microRNAs due to a phenomena called the Hill-Robertson interference, or HRI (58, 59). Both positive and purifying selection results in HRI, the former by selective sweeps (60) and the latter by background selection (61). This type of interference between linked loci has been used to explain the quantitatively reduced impact of selection compared to non-adaptive forces across whole genomes (62), and it is likely to account for the evolutionary pattern of tightly linked sequences such as clustered microRNAs.

We propose an evolutionary model for the origin and evolution of microRNA clusters which we call the 'drift-draft' model. New microRNA hairpins often emerge *de novo* in an existing transcript (44, 63). Under our model of microRNA evolution we envision two scenarios. First, the new microRNA appears within a primary microRNA transcript, therefore both microRNAs will be tightly linked in the genome. The older microRNA is subject to strong purifying selection so that the new microRNA is (almost) invisible to natural selection due to HRI as recombination between the two microRNAs is virtually absent. In a second scenario, a novel microRNA may appear and provide selective advantage to the host genome. Due to HRI, positive selection will drive the evolution of the novel microRNA while, again, non-adaptive forces would dominate the evolutionary fate of the other microRNAs in the cluster. Our drift-draft model is consistent with the observations that most clusters contain members of only a few families, that clusters are relatively young, and that they evolve as a single unit. It also explains why tandem duplication may happen within pre-formed clusters: changes in the number of microRNAs linked to a selectively constrained neighbor will have a minor impact on the function of the cluster. Future development of theoretical models and analysis of population polymorphism data will explore the validity of this model.

In the light of our observations, the emergence of polycistronic microRNAs is largely non-adaptive, and the maintenance of the clusters is most likely a by-product of tight genomic linkage. However, a potential role of natural selection in functional diversification of clusters is yet to be elucidated. The linkage of microRNAs to other loci (microRNAs or other genes) has been so far ignored in microRNA evolutionary studies. The impact of genomic linkage has been shown to be a crucial factor in the evolution of protein coding genes, but may be even more important in the evolution of microRNAs and other small RNA coding loci.

## SUPPLEMENTARY DATA STATEMENT

Supplementary Data are available at NAR online: Supplementary tables 1-2, Supplementary figure 1 and Supplementary datasets 1-2.

## FUNDING

This work was supported by the Wellcome Trust Institutional Strategic Support Fund (097820/Z/11/Z) and the Biotechnology and Biological Sciences Research Council (BB/G011346/1 and BB/H017801/1). MN is funded by a Wellcome Trust PhD studentship.

## ACKNOWLEDGEMENTS

We thank Casey Bergman for helpful discussion.

# FIGURE LEGENDS

**Figure 1. Mechanisms of microRNA cluster emergence.** (A) Put together: microRNAs in different genomic loci involved in related functional pathways end up being clustered in the genome. (B) Left together: microRNAs in different genomic loci become clustered in the genome as a by-product of genome rearrangements. (C) Tandem duplication: a microRNA is duplicated in tandem producing a polycistronic transcript. (D) New hairpin: a novel microRNA emerges within the primary transcript of an existing microRNA.

**Figure 2. Clusters of microRNAs in the *D. melenogaster* genome.** (A) Box-plots of expression correlation (Pearson) between pairs of neighboring microRNAs as a function of the genomic distance. (B) Frequency distribution of the number of different microRNA families in each cluster (black boxes) and the number of microRNAs per cluster (white boxes). (C) Bubble-plot of microRNA cluster sizes against the number of families. The number in each bubble is the number of instances of clusters of a given size (*y*-axis) with a given number of families (*x*-axis).

**Figure 3. Origin of *D. melanogaster* microRNA clusters.** Clusters emerging in a given lineage are listed on the corresponding branch of the evolutionary tree. Clusters that formed by the emergence of new hairpins in existing transcripts are labelled with a [n], and clusters formed by tandem duplication with a [d]. The label [u] indicates that we cannot infer whether the cluster originally came from a tandem duplication or a new hairpin formation. For clusters with more than two members only the first and last microRNA are shown separated by a tilde.

**Table 1. Origin of *Drosophila melanogaster* microRNA clusters**

| Cluster | Source | Lineage | Notes |
|---|---|---|---|
| 999/4969 | new hairpin | Melanogaster | Original miRNA: mir-999 |
| 982/303/983-1/983-2/984 | new hairpin | Melanogaster | Multiple emergence within a conserved gene |
| 969/210 | new hairpin | *Drosophila* | Original microRNA: mir-210 |
| 124/287 | new hairpin | *Drosophila* | Original microRNA: mir-124 |
| 972/973/974/2499/4966/975/976/977/978/979 | new hairpin | *Drosophila* | |
| 959/960/961/962/963/964 | new hairpin | *Drosophila* | |
| 1002/968 | new hairpin | *Drosophila* | |
| 281-2/281-1 | duplication | *Drosophila* | |
| 310/311/312/313/2498/991/992 | duplication | *Drosophila* | Probably two clusters: 310/311/312/313 and 2498/991/992 |
| 6-3/6-2/6-1/5/4/286/3/309 | new hairpin | Insects | Cluster may be older (see main text) |
| 998/11 | new hairpin | Insects | |
| 994/318 | new hairpin | Insects | |
| 279/996 | duplication | Insects | |
| 9c/306/79/9b | unknown | Insects | |
| 283/304/12 | new hairpin | Protostomes | |
| 275/305 | new hairpin | Protostomes | |
| 317/277/34 | new hairpin | Protostomes | Original microRNA: mir-34 |
| 13b-1/13a/2c | duplication | Protostomes | The original mir-2 cluster probably emerged by *de novo* acquisition of mir-2 nearby mir-71 (see main text) |
| 2a-2/2a-1/2b-2 | duplication | Protostomes | " |
| 92a/92b | duplication | Metazoans | Duplications in insects and chordates may be independent |
| 100/let-7/125 | unknown | Metazoans | mir-100 and mir-125 are paralogs |

**Table 2. Non-clustered *Drosophila* microRNAs that are clustered in other species**

| microRNA | Clustered homolog* | Cluster source |
|---|---|---|
| mir-1/mir-133 | Clustered together in animals. >10kb in *D. melanogaster* | new hairpin |
| mir-7 | Clustered with mir-1179 in mammals | new hairpin |
| " | Clustered with mir-3529 in *Gallus gallus* | new hairpin |
| " | Clustered with mir-1720 in *Gallus gallus* | new hairpin |
| mir-8 | Tandem copies in chordates (mir-200) | duplication |
| mir-10 | Clustered with mir-2886 in *Bos taurus* | new hairpin |
| " | Clustered with mir-1713 in *Gallus gallus* | new hairpin |
| mir-31a | Tandem duplication in *Rattus norvegicus* | duplication |
| " | Tandem duplication in *Schmidtea mediterranea* | duplication |
| mir-33 | Tandem duplication in *Branchiostoma floridae* | duplication |
| mir-87 | Tandem duplication in insects. One copy lost in *Drosophila* | duplication |
| mir-137 | Clustered with mir-2682 in *Homo sapiens* | new hairpin |
| mir-184 | Tandem duplication in *Capitella teleta* | duplication |
| mir-193 | Clustered with mir-365 in vertebrates | new hairpin |
| mir-219 | Clustered with mir-2964 in vertebrates | new hairpin |
| mir-252 | Tandem duplication in *Acyrthosiphon pisum* | duplication |
| " | Tandem duplication and novel mir-2001 in *Lottia gigantea* and *Capitella teleta* | duplication/new hairpin |
| mir-263a/b | Clustered together in *Daphnia pulex*. Not clustered in other insects | duplication |
| mir-276a/b | Clustered together in *Drosophila* lineage. >10kb in *D. melanogaster* | duplication |
| mir-285 | Tandem duplication in vertebrates | duplication |
| " | Clustered with mir-3556 and mir-3587 in *Rattus norvegicus* | new hairpin |

*As annotated in miRBase (http://mirbase.org)

**Figure 1**

(A) put together

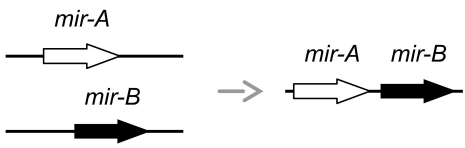

(B) left together

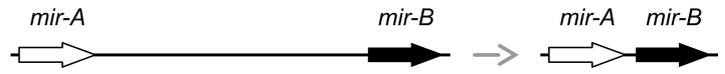

(C) tandem duplication

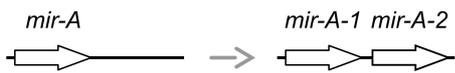

(D) new hairpin

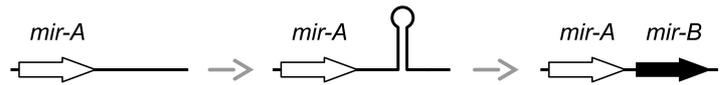

**Figure 2**

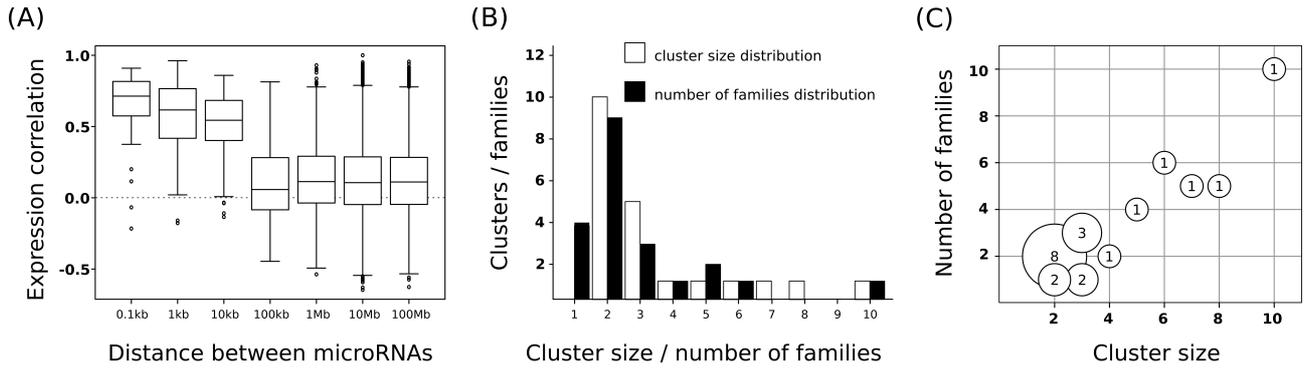

**Figure 3**

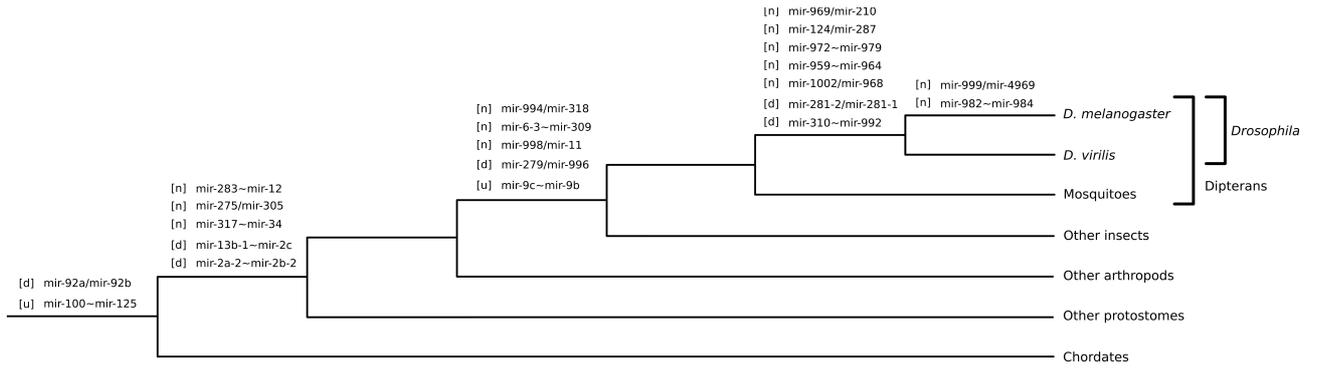